\documentclass[prl,twocolumn,letterpaper,aps]{revtex4-1}

\usepackage{amsmath}
\usepackage{amssymb}
\usepackage{amsfonts}
\usepackage{graphicx}
\usepackage{dcolumn}
\usepackage{float}
\usepackage{bm}
\usepackage{mathrsfs}
\usepackage{tabularx}
\usepackage{bigstrut}
\usepackage{xfrac}
\usepackage{listings}
\usepackage[usenames,dvipsnames]{xcolor}
\usepackage[hyperindex, breaklinks,colorlinks = true,linkcolor = MidnightBlue,urlcolor=blue,citecolor=blue]{hyperref}

\begin{document}

\title{ Quantum Interference Control of Photocurrents in Semiconductors by Nonlinear Optical Absorption Processes }

\author{Kai Wang$^{1}$, Rodrigo A. Muniz$^{2}$, J. E. Sipe$^{2}$, and S. T. Cundiff$^{1}$}

\affiliation{$^{1}$Physics Department, University of Michigan, Ann Arbor, Michigan
48109, USA \\
 $^{2}$Department of Physics, University of Toronto, Toronto, Ontario M5S 1A7, Canada }

\date{\today}

\begin{abstract}
We report experiments demonstrating Quantum Interference Control (QuIC) based on two nonlinear optical absorption processes in semiconductors. We use two optical beams of frequencies $\omega$ and $3\omega /2$ incident on AlGaAs, and measure the injection current due to the interference between 2- and 3-photon absorption processes. We analyze the dependence of the injection current on the intensities and phases of the incident fields.
\end{abstract}

\maketitle

When different quantum evolution pathways can lead a system from the same initial to a final state, quantum interference between the routes leads to enhancement or suppression of the transition rate.
Quantum interference control (QuIC) based on optical processes of different photon numbers has been used for exciting and controlling target states in both molecular and crystalline systems.
It has been used for molecular excitation and ionization \cite{Manykin,Shapiro03,Brumer86,Chen90,Zhu95, Nagai06},
and in semiconductors it was first used for asymmetric photoejection
\cite{Kurizki89,Baranova90,Lawandy90, Baranova91,Dupont95} and later
for current injection \cite{Atanasov96,Hache97}. Quantum interference of optical absorption processes can be constructive in some regions of the Brillouin zone (BZ) while destructive in others, which
results into a net overall current control. 
To date every use of QuIC in crystalline materials has involved 1- and 2-photon absorption processes; 
1+2 QuIC has been used for charge \cite{Atanasov96, Hache97,Rioux12} and spin \cite{Bhat00, Stevens02, Stevens03, Hubner03, Zhao06} current injection in semiconductors, as well as current injection in graphene \cite{Sun10,Rioux11, Rao12}, topological insulators \cite{Muniz14,Bas15, Bas16}, and transition metal dichalcogenides \cite{Muniz15, Cui15}. 
It has also been theoretically investigated for current injection in graphene nanoribbons \cite{Salazar16}, spin currents in topological insulators \cite{Muniz14}, and spin and valley currents in transition metal dichalcogenides \cite{Muniz15}.

Second and third harmonic generation are the typical techniques used to obtain optical fields with the frequencies appropriate for 1+2 and 1+3 QuIC, respectively \cite{Rioux12,Shapiro03}. 
And second harmonic generation has also been used for 2+4 QuIC of atomic ionization \cite{Papastathopoulos99}. 
The generation of optical fields of frequencies appropriate for $m+n$ QuIC with fractional ratio $n/m$ is a much harder task, which has limited the study of QuIC based on such processes. One way to study more general $n+m$ QuIC is to use phase-coherent frequency combs \cite{Cundiff03},  especially for fractional ratios $n/m <2$ as these cases do not require the frequency range of the comb to be too broad.
The use optical frequency combs for QuIC experiments presents  an opportunity to separately study several nonlinear optical processes in semiconductors. We note that when considering these processes in the frequency domain, all possible combinations of comb lines must be considered, not just harmonics of individual lines, as is already the case in considering ``second'' harmonic generation where not considering sum-frequency terms between comb-lines leads to the incorrect conclusion that the second harmonic pulse train has twice the repetition rate of the fundamental \cite{Cundiff_2002}.

In this letter we report experiments demonstrating QuIC of photocurrents using 2- and 3-photon absorption (2PA and 3PA) processes in AlGaAs. 
A theoretical study of 2+3 QuIC of photocurrents in AlGaAs is presented in another paper \cite{Muniz17}.  
In crystalline materials, QuIC involving higher order processes lead to higher swarm velocities due to better localization of carriers  \cite{Muniz17} in the BZ.  Thus $m+n$ QuIC experiments in semiconductors not only allow the study of some nonlinear optical processes separately, but they also open the possibility of precise-probing properties of the electronic states in regions of the BZ.

\begin{figure}[htb!]
\includegraphics[width=0.99\columnwidth]{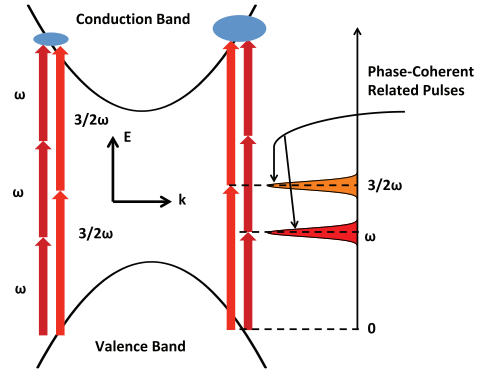} 
\caption[]{ Illustration of the quantum interference between the 2PA and 3PA pathways for the excitation of an electron from a valence to a conduction band. If the two pulses are phase coherent related, the QuIC current is generated due to the quantum interference between them.  }
\label{fig:bands}
\end{figure}

\begin{figure}[htb!]
	\includegraphics[width=0.99\columnwidth]{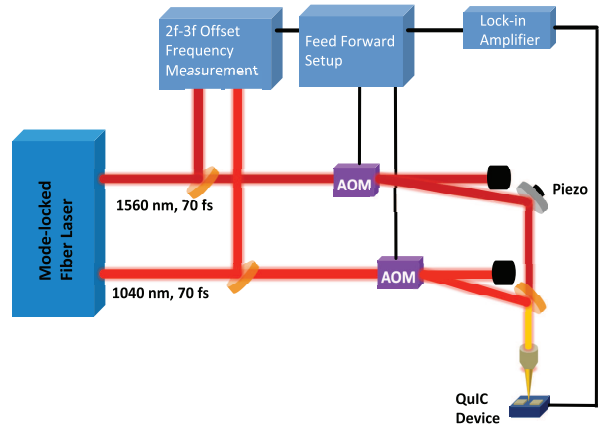} 
\caption[]{ The experimental setup is shown schematically. The comb offset frequency is measured using a 2f-3f interferometer and the beat note signal is then amplified to drive the AOM inserted in the beam path. The diffracted beam of order $-1$ is the offset-frequency stabilized beam used in the QuIC measurement. 
The two diffracted beams are then incident on a metal-semiconductor-metal (MSM) device. 
The signal is detected through a lock-in amplifier referenced by the offset frequency set by feed-forward setup. }
\label{fig:setup}
\end{figure}
\begin{figure}[htb!]
\includegraphics[width=0.99\columnwidth]{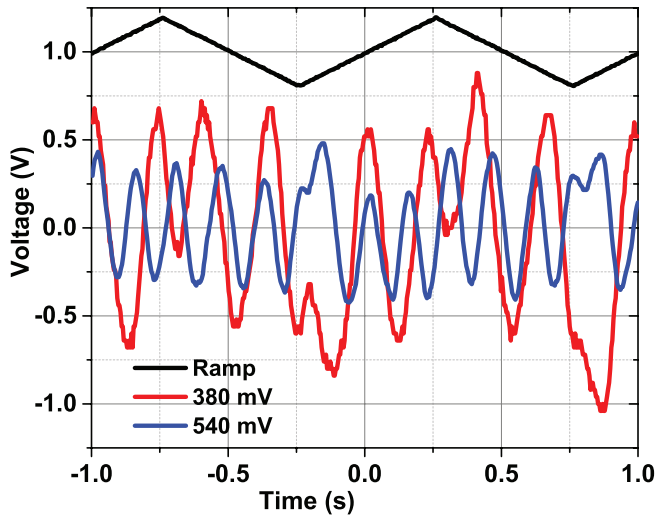} 
\caption[]{Phase dependence of the 2+3 QuIC current.
The fringes are associated with the phase ramping of the $1560\,$nm beam. The black curve is the ramping voltage applied on the piezo to change the phase of $1560\,$nm beam. When increasing the amplitude of the ramping voltage by a factor of 1.4, from 380 mV (red) to 540 mV (blue), the number of fringes is increased to about 1.5 times, approximately the same ratio. The estimated displacement of piezo is about $1.54\pm0.05$nm/mV.}
\label{fig:phase} 
\end{figure}
\begin{figure}[htb!]
\includegraphics[width=0.99\columnwidth]{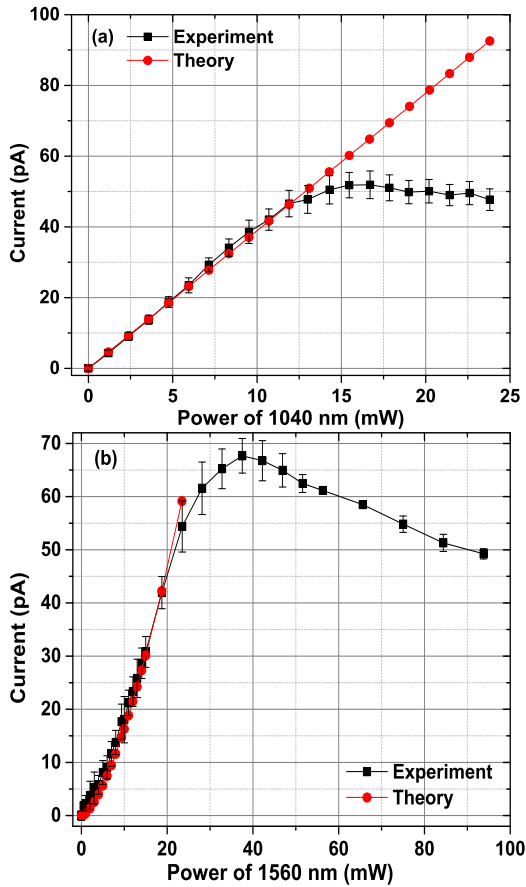} 
\caption[]{ (a) Amplitude of the 2+3 QuIC in terms of the  average power of the $1040\,$nm beam ($3\omega/2$ field) when the power of the $1560\,$nm beam($\omega$ field) is fixed at $20\,$mW.
The red circles indicate the expected dependence based on Eq. \eqref{eq:J23}. 
(b) Amplitude of the 2+3 QuIC in terms of the average power of the $1560\,$nm beam when the power of the $1040\,$nm beam is fixed at $8.5\,$mW. The red circles indicate the expected dependence based on Eq. \eqref{eq:J23}. }
\label{fig:power}
\end{figure}

We denote the two incident optical field amplitudes by $\boldsymbol{E}_{\omega}=E_{\omega}e^{i\phi_{\omega}}\hat{\boldsymbol{e}}_{\omega}$
and $\boldsymbol{E}_{3\omega/2}=E_{3\omega/2}e^{i\phi_{3\omega/2}}\hat{\boldsymbol{e}}_{3\omega/2}$,
where $E_{\omega}>0$ and $E_{3\omega/2}>0$ are the field magnitudes, the unit vectors $\hat{\boldsymbol{e}}_{\omega}$ and $\hat{\boldsymbol{e}}_{3\omega/2}$
indicate the field polarizations, and $\phi_{\omega}$ and $\phi_{3\omega/2}$
indicate the field phases. When either the $\omega$ or the $3\omega/2$
field alone is incident on the sample, the distribution of injected carriers in the BZ is nonpolar, so there is no net current.
However, when both fields are incident, the quantum interference between 3PA with frequency
$\omega$ and 2PA with frequency $3\omega/2$ leads to a polar distribution of carriers in the BZ, as sketched in Fig.~\ref{fig:bands}.
The injection rate of the current density $J$ due to 2+3 QuIC can be written in terms of a tensor $\eta_{2+3}\left(\Omega\right)$ such that
\begin{equation}
\frac{d}{dt}J_{2+3}^{a}=\eta_{2+3}^{abcdef}\left(\Omega\right)e_{\omega}^{b\ast}e_{\omega}^{c\ast}e_{\omega}^{d\ast}e_{3\omega/2}^{e}e_{3\omega/2}^{f} e^{i\Delta\phi}E_{\omega}^{3}E_{3\omega/2}^{2}+c.c.,\label{eq:J23}
\end{equation}
 where $\hbar\Omega=3\hbar\omega$ is the total photon energy of the
absorption processes, and we define the relative phase parameter $\Delta\phi=2\phi_{3\omega/2}-3\phi_{\omega}$ \cite{Muniz17}. 
If the $\omega$ and $3\omega/2$ fields correspond to pulse trains from the same optical frequency comb, or with the same repetition rate, the QuIC current is modulated by the offset frequency $f_{0}$ through the phase parameter $\Delta\phi = 2\pi f_{0}t$. Thereby, we could characterize the offset frequency of a frequency comb based on the QuIC current \cite{Fortier04}.
The lattice symmetries impose constraints on the components of the
tensor $\eta_{2+3}\left(\Omega\right)$. For most frequencies, the
largest independent component \cite{Muniz17} for AlGaAs is $\eta_{2+3}^{xxxxxx}\left(\Omega\right)$,
which corresponds to the polarization of both fields and the current being along the $[100]$ crystal axis. 
Unless specified otherwise, these are the polarizations used in all of our discussion.
We measure the injection current with electrodes aligned along the$ [100]$
 crystal axis. 
We illustrate our experimental setup in Fig.~\ref{fig:setup}, and we present the observed dependence of the detected signal on the intensities and phases of the fields in Figs.~\ref{fig:phase} and \ref{fig:power} respectively.


The experimental setup is shown in Fig.~\ref{fig:setup}. A custom laser system (MenloSystems) outputs two femtosecond pulse trains derived from a common oscillator at different wavelengths:
one is $400\,$mW and centered at $1560\,$nm and the other
is $740\,$ mW and centered at $1040\,$nm. Both beams have
a pulse duration of about $70\,$fs and repetition rate of $250.583\,$MHz. The $1560\,$nm and $1040\,$nm beams are also respectively referred to as $\omega$ and $3\omega/2$ fields in this paper.
We measure the offset frequency of the laser comb using the heterodyne beat note produced in a $2f$-$3f$ self-referencing interferometer,
for which we double the frequency of $1040\,$nm beam with a BBO crystal,
and triple the $1560\,$nm beam with a periodically poled lithium niobate (PPLN) crystal; although the PPLN is designed for second harmonic generation, it also produces weak third harmonic. 
The beat note measured by the detector is then used as a source in a feed-forward setup to stabilize the offset frequency. 
In the feed-forward setup \cite{koke10}, an acousto-optic modulator (AOM) is inserted in the beam path driven by the amplified beat note, so the diffracted beam of order $-1$ is the resulting offset-frequency-stabilized beam.  
The feed-forward setup also allows us to control the offset frequency. 
Details are given in the supplementary material. In the experiment, in order to avoid the spurious signal induced by a harmonic of the AOM driving RF, the offset frequencies of the beams are $f_{1560}^{off}=20\,$KHz and $f_{1040}^{off}=14\,$KHz, and the measured beat note has frequency $3f_{1560}^{off}-2f_{1040}^{off}= 32\,$KHz.
The QuIC current is then detected by a lock-in amplifier.

The sample device is made from epitaxially grown AlGaAs on a GaAs substrate. The bandgaps of the AlGaAs layers correspond to wavelengths
lower than $700\,$nm, so the linear absorption of $1040\,$nm beam
and 2PA of the $1560\,$nm beam are suppressed. 
We detect the 2+3 QuIC current with two electrodes made from annealed Au/Ge, which
forms an Ohmic contact between the metal and semiconductor. 
The two beams are incident on the same spot between the electrodes, which are separated by around
$20\,\mu$m. The spot radius of the $1560\,$nm beam on the device is approximately
$2\,\mu$m, while that of the $1040\,$nm beam is approximately
$3\,\mu$m. The electrodes are aligned such that they measure the
photocurrent along the $[100]$ crystal axis, which is also referred to as the horizontal direction, and the sample is electronically unbiased for the 2+3 QuIC measurements. Dispersion will limit the depth over which the QuIC process contributes to the measured signal due the phase slip between the $2\omega$ and $3\omega$ beams. Simulations estimate this depth to be approximately 2.36 $\mu$m.

The first feature of the 2+3 QuIC current we consider is its phase dependence. Both beams are setup to be horizontally polarized. Using piezoelectric transducers, the relative phase between the two beams is slowly ramped over several fringes. 
The photocurrent is detected with the assistance of a lock-in amplifier. We capture time traces of the lock-in amplifier output with an oscilloscope, shown in Fig. \ref{fig:phase}.

Next we examine the dependence of the 2+3 QuIC signal on the intensity of the fields.
Since the QuIC signal is measured in a sensitive interferometric setup, environmental vibrations lead to amplitude and phase fluctuations of the current even when the field intensities are constant. 
When using a lock-in amplifier to record data, there is also phase drift in the measurement. 
Therefore, in order to determine the current magnitude $r$ statistically, we measure the QuIC current through two outputs of the lock-in amplifier (in-phase component as $X$, and quadrature component as $Y$), and calculate the signal magnitude, $r=\sqrt{X^{2}+Y^{2}}$, to remove the phase dependency. 
Using the current magnitude $r$, we examine the dependence of the 2+3 QuIC signal on the intensity
of the fields. 
The beam spot sizes and the repetition rates are kept
constant and both beams are horizontally polarized, so we control the intensities only by the power of lasers. 
We use a rotary variable attenuator inserted in the beam path to adjust the average optical power. 
First we fix the power of the $\omega$ field at $20\,$mW while we vary the power of the $3\omega/2$ field.
We show the measured amplitude of the 2+3 QuIC signal in this situation in Fig. \ref{fig:power}(a).
From the plot, we can first confirm that
the signal indeed corresponds to an interference process as it vanishes when the $3\omega/2$ field is not incident (zero power).
We also confirm that the signal magnitude depends linearly on the power of the $3\omega/2$ field for low powers, which is in agreement with Eq.~\eqref{eq:J23}.
We then fix the power of the $3\omega/2$ field at $8.5\,$mW while we vary the power of the $\omega$ field. 
We show the measured amplitude of the 2+3 QuIC signal in this situation in Fig.~\ref{fig:power}(b).
For low powers of the $\omega$ field, the magnitude of the QuIC signal can be fitted by a power-law with exponent $3/2$ in terms of the $\omega$ field power, which again is in agreement with Eq.~\eqref{eq:J23}.
For high enough laser powers, the 2+3 QuIC signal is expected to saturate due to Pauli blocking of the electronic transition from the valence to the conduction band, and the injection of other carriers through higher order optical process, as they can at least partially shield the injection current. 
For the fixed values of the powers of the beams we considered, the saturation thresholds are around $12\,$mW for the $3\omega/2$ field, and around $25\,$mW for the $\omega$ field, which are both in good agreement with theoretical estimates \cite{Muniz17}. The highest current amplitude observed in the experiments is around $70\,$pA.

\begin{figure}[htb!]
\includegraphics[width=0.99\columnwidth]{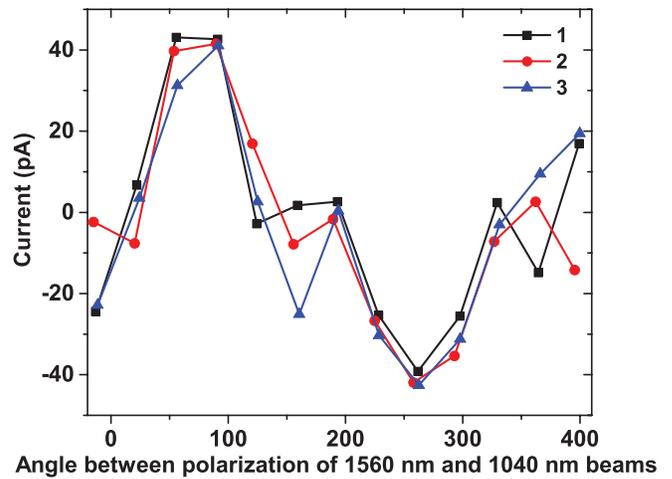} \protect\caption[]{ The QuIC signal as a function of angle between directions of polarization of the $3\omega/2$ ($1040\,$nm) and $\omega$ ($1560\,$nm) fields. 
In the measurement, the polarization of the 3$\omega$/2 ($1040\,$nm) field is horizontal, which is along the $[100]$ crystal axis and the direction is labeled x. The polarization of  $\omega$ ($1560\,$nm) is rotated by half wave plate. The same measurement is repeated three times and it shows a sinusoidal-like behavior as expected from Eq.~\eqref{eq:polariz} (Black).}
\label{fig:hwplate}
\end{figure}

To analyze the dependence of the 2+3 QuIC signal on the polarizations of the fields, we use a polarizer in the beam path to ensure that the incident light is linearly polarized.  After the polarizer, we insert a half-wave plate to vary the polarization direction. 
When the polarization of $3\omega/2$ field is horizontal ($\hat{\bm{e}}_{3\omega/2} = \hat{\bm{x}}$) and we vary the polarization of the $\omega$ field ($\hat{\bm{e}}_{\omega} =\hat{\bm{\theta}}$) eqn.~\ref{eq:J23} becomes
\begin{align}
\frac{d}{dt}J_{2+3}^{x} = & \cos\theta \left[\eta_{2+3}^{xxxxxx} \cos^{2}\theta +\eta_{2+3}^{xxyyxx} \sin^{2} \theta \right] \nonumber \\
& \> \times  e^{i\Delta\phi}E_{\omega}^{3}E_{3\omega/2}^{2} +c.c.,
\label{eq:polariz}
\end{align}
where $x$ and $y$ respectively denote the horizontal and vertical directions, and $\theta$ is the angle between the direction of polarization of the $\omega$ and $3\omega/2$ fields. 
To test the relation between the QuIC current and the polarizations of beams, we rotate the half wave plate of the $\omega$ field at a speed faster than the phase drift rate, and record the in-phase component of lock-in amplifier at around every $18^\circ$ of the half-wave plate. 
In Fig.~\ref{fig:hwplate} we plot the QuIC current dependence on $\theta$, as the polarization of the $3\omega/2$ beam is fixed at the horizontal direction. 
It shows a sinusoidal-like behavior as expected according to Eq.~\eqref{eq:polariz}, which reaffirms that the signal measured indeed corresponds to the 2+3 QuIC current. 
The average powers of the two fields are fixed at $12\,$mW for the $3\omega/2$ field and $40\,$mW for the $\omega$ field.

According to theoretical predictions \cite{Muniz17}, there should be no horizontal 2+3 QuIC current when the $\omega$ and $3\omega/2$ fields are both vertically polarized, as well as when the $\omega$ field is horizontal and the $3\omega/2$ field is vertical. 
However, we detect a signal when both beams are vertically polarized, and when the $\omega$ field is horizontally polarized and $3\omega/2$ field is vertically polarized, although those signals are much smaller than when both fields are horizontally polarized. 
We attribute these results to residual depletion DC fields of the device, as a depletion field leads to field-induced QuIC current along the horizontal direction \cite{Wahlstrand11}. This effect is not included in the theoretical analysis leading to eqn.~\ref{eq:polariz}, thus it predicts no signal in this case. Our current detection apparatus can not distinguish the 2+3 QuIC current from the field-induced QuIC current.

In conclusion, our results demonstrate 2+3 QuIC of photocurrents in semiconductors.  
QuIC based on nonlinear processes can be used for carrier injection strongly localized in the BZ, which allows the detailed study of carrier dynamics in semiconductors. 
The phase dependence of the QuIC photocurrent has been used as a means to measure phase parameters of optical fields by detecting the currents they inject in a sample \cite{Roos05optlet,Roos05laser,Smith07}.  This strategy has been used for the characterization \cite{Roos04,Fortier04} and stabilization \cite{Roos05optlet, Roos05laser, Roos05josaB} of optical frequency combs. 
Thus 2+3 QuIC of photocurrents is not only enabled by frequency combs, but it also has a natural application in the  stabilization of combs that do not have an octave-spanning frequency range.
Moreover, since only nonlinear optical processes are involved in 2+3 QuIC, the fields have a weak power law attenuation as they propagate through the absorbing material, instead of the exponential attenuation of linear absorption.
Thus it is possible to use a waveguide structure for the absorption region of a device in order to increase the signal-to-noise ratio. 
Such a scenario allows for easy integration with devices on-chip, provided the issues of phase and mode matching are addressed, so we expect that one of the most immediate applications of 2+3 QuIC could be the on-chip stabilization of optical frequency combs that are not octave-spanning.


This work has been supported by DARPA through the DODOS program.

\bibliography{bibliQuIC}
\bibliographystyle{apsrev4-1}
\end{document}